\documentclass[aps,prl,twocolumn,superscriptaddress]{revtex4-2}

\usepackage{amsmath}
\usepackage{amssymb}
\usepackage{graphicx}
\usepackage{color}
\usepackage[colorlinks,citecolor=blue,linkcolor=blue,urlcolor=blue]{hyperref}

\newcounter{fnnumber}

\newcommand{\dd}{\mathrm{d}}

\newcommand{\frd}[2]{\frac{\mathrm{d}#2}{\mathrm{d}#1}}

\newcommand{\mbb}[1]{\mathbb{#1}}
\newcommand{\mbf}[1]{\mathbf{#1}}
\newcommand{\mcal}[1]{\mathcal{#1}}
\newcommand{\mr}[1]{\mathrm{#1}}
\newcommand{\pd}{\partial}

\begin{document}
\title{Emergent Symmetry and Phase Transitions on the Domain Wall of $\mathbb{Z}_{2}$ Topological Orders}

\author{Hong-Hao Song}
\thanks{These authors contributed equally.}
\affiliation{Kavli Institute for Theoretical Sciences and CAS Center for Excellence in Topological Quantum Computation, University of Chinese Academy of Sciences, Beijing 100190, China}

\author{Chen Peng}
\thanks{These authors contributed equally.}
\affiliation{Kavli Institute for Theoretical Sciences and CAS Center for Excellence in Topological Quantum Computation, University of Chinese Academy of Sciences, Beijing 100190, China}

\author{Rui-Zhen Huang}
\affiliation{Graduate School of China Academy of Engineering Physics, Beijing 100193, China}

\author{Long Zhang}
\email{longzhang@ucas.ac.cn}
\affiliation{Kavli Institute for Theoretical Sciences and CAS Center for Excellence in Topological Quantum Computation, University of Chinese Academy of Sciences, Beijing 100190, China}

\date{\today}

\begin{abstract}
The one-dimensional (1D) domain wall of 2D $\mathbb{Z}_{2}$ topological orders is studied theoretically. The Ising domain wall model is shown to have an emergent SU(2)$_{1}$ conformal symmetry because of a hidden nonsymmorphic octahedral symmetry. While a weak magnetic field is an irrelevant perturbation to the bulk topological orders, it induces a domain wall transition from the Tomonaga-Luttinger liquid to a ferromagnetic order, which spontaneously breaks the anomalous $\mathbb{Z}_{2}$ symmetry and the time-reversal symmetry on the domain wall. Moreover, the gapless domain wall state also realizes a 1D topological quantum critical point between a $\mathbb{Z}_{2}^{T}$-symmetry-protected topological phase and a trivial phase, thus demonstrating the holographic construction of topological transitions.
\end{abstract}

\maketitle 

{\it Introduction.---} Topological phases often have nontrivial gapless boundary states, which bear valuable information of the bulk topological states via the bulk-boundary correspondence. The chiral edge states of quantum Hall insulators (QHIs)~\cite{Halperin1982, Wen1991a, Wen1990b} and the helical edge states of quantum spin Hall insulators (QSHIs)~\cite{Kane2005, Bernevig2006} are prominent examples. The chiral edge states of QHIs are robust against generic local perturbations as long as the bulk energy gap is not closed, and are characterized by a nonzero thermal Hall conductance~\cite{Kane1997, Kitaev2006a}. In contrast, the helical edge states of QSHIs are protected only when the time-reversal symmetry~\cite{Kane2005a, Wu2006, Xu2006} or the spin rotation symmetry~\cite{Sheng2006} is preserved, which prohibits the backscattering of electrons at the edge.

Nonchiral topological phases with vanishing thermal Hall conductance may have gapped boundary states, where a group of bosonic topological excitations condense at the boundary and render all other topological excitations confined because of their nontrivial mutual statistics with the condensed ones~\cite{Kapustin2011, Beigi2011, Kitaev2012, Levin2013, Barkeshli2013b, Barkeshli2013c}. Gapped domain walls between topological orders can also be constructed if topological excitations can tunnel through the domain wall in a way compatible with their braiding statistics in the bulk topological orders~\cite{Lan2015}.

Gapless domain walls of nonchiral topological orders are more elusive. Recently, a systematic construction of one-dimensional (1D) gapless domain walls of 2D topological orders was proposed~\cite{Bao2022}. In particular, for the domain wall between $\mbb{Z}_{2}$ topological orders, i.e., the toric code~\cite{Kitaev2003, Levin2005} and the double semion models~\cite{Levin2005}, the low-energy spectrum was calculated numerically and found to be consistent with the SU(2)$_{1}$ Wess-Zumino-Witten (WZW) conformal field theory~\cite{Bao2022}, even though the domain wall model does not have any continuous symmetry. An argument based on the anyon condensation scenario was proposed for the emergent SU(2)$_{1}$ conformal symmetry~\cite{Bao2022}, but a microscopic understanding is still lacking. In addition, it is also desirable to find out the general domain wall phase diagram and particularly the possible transitions to gapped domain wall states.

In this work, we study the emergent symmetry and the phase diagram on the domain wall of $\mbb{Z}_{2}$ topological orders. We find an exact unitary transformation that maps the domain wall model into a deformed Heisenberg chain and thus identify a hidden nonsymmorphic octahedral symmetry. The low-energy effective theory is given by the SU(2)$_{1}$ WZW model, because all relevant perturbations are prohibited by the nonsymmorphic octahedral symmetry. We then consider a more generic domain wall perturbed by a nonzero magnetic field of the bulk topological orders, which induces an Ising interaction on the domain wall model. It drives the domain wall transitions from the gapless Tomonaga-Luttinger liquid (TLL) into gapped phases with ferromagnetic (FM) or antiferromagnetic (AF) orders, which spontaneously break the time-reversal symmetry and the anomalous $\mbb{Z}_{2}$ symmetry inherited from the bulk $\mbb{Z}_{2}$ gauge structure. Moreover, the gapless domain wall state corresponds to a 1D quantum critical point (QCP) between a $\mbb{Z}_{2}^{T}$-symmetry-protected topological (SPT) phase and a trivial phase, hence it is an example of the holographic construction of topological QCPs as gapless domain walls of one-higher-dimensional topological phases~\cite{Chen2013g, Tsui2015a}.

{\it Emergent SU(2)$_{1}$ conformal symmetry.---} The 1D domain wall between the toric code and the double semion topological orders is described by the Ising domain wall model (IDWM)~\cite{Bao2022},
\begin{equation}
H_{0}=-\sum_{l}\sigma^{z}_{l-1}\sigma^{x}_{l}\sigma^{z}_{l+1}-\sum_{l}\sigma^{y}_{l},
\label{eq:idw}
\end{equation}
where $\sigma^{x}_{l}$, $\sigma^{y}_{l}$ and $\sigma^{z}_{l}$ are the Pauli operators on site $l$. The IDWM has an anomalous $\mbb{Z}_{2}$ self-dual symmetry $\mcal{S}$, which is inherited from the bulk $\mbb{Z}_{2}$ gauge structure~\cite{Levin2012, Bao2022},
\begin{equation}
\mcal{S}=\prod_{l}\sigma^{x}_{l}\prod_{l}\exp\Big(\frac{i\pi}{4}(\sigma^{z}_{l}\sigma^{z}_{l+1}-\sigma^{z}_{l}-1)\Big),
\label{eq:selfdual}
\end{equation}
which exchanges the two terms in Eq.~(\ref{eq:idw}). $H_{0}$ also has an antiunitary $\mbb{Z}_{2}^{T}$ symmetry generated by $\mcal{A}=(\prod_{l}\sigma^{z}_{l})\mcal{T}$, where $\mcal{T}$ is the usual time-reversal transformation.

Despite the fairly low symmetry of the IDWM, numerical calculations suggested that its low-energy spectrum closely resembles that of the spin-$1/2$ Heisenberg chain and is consistent with the SU(2)$_{1}$ WZW theory~\cite{Bao2022}. Further numerical evidence supporting the SU(2)$_{1}$ WZW effective theory is provided in the Supplemental Materials (SM)~\footnote{See the Supplemental Materials for details, which include Refs.~\cite{Bao2022, Yang2025, Affleck1986, Gogolin2004, Senechal2004, Calabrese2004, Furukawa2009, Calabrese2009c, Affleck1998a}.}\setcounter{fnnumber}{\thefootnote} based on the two-interval mutual information of the ground state~\cite{Furukawa2009, Calabrese2009c}. However, the microscopic origin of the emergent SU(2)$_{1}$ conformal symmetry was not clear.

\begin{figure}[tb]
\centering
\includegraphics[width=\columnwidth]{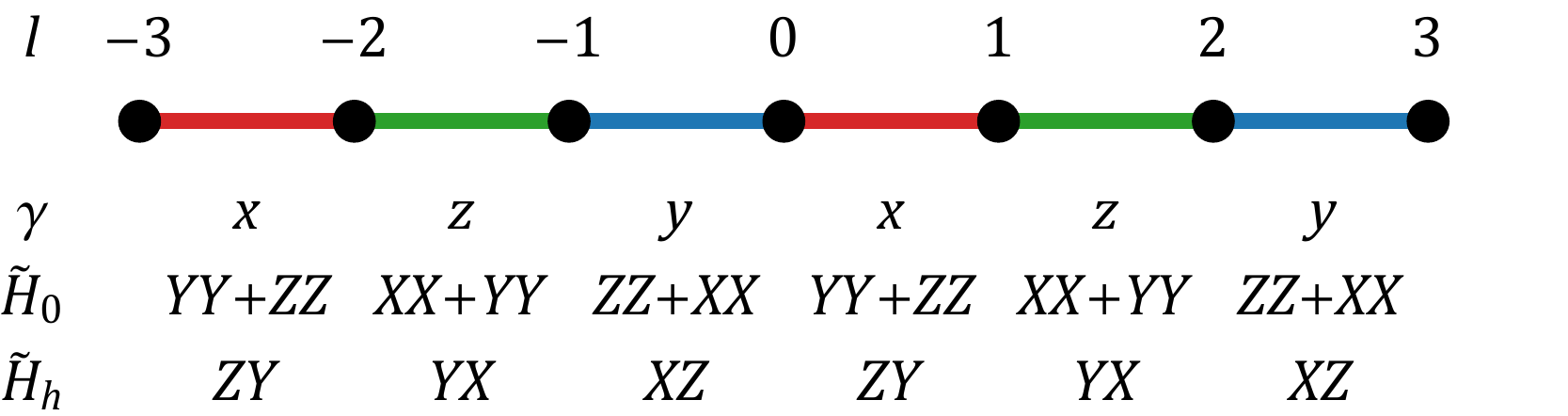}
\caption{Schematic illustration of the bond and spin operator indices in $\tilde{H}_{0}$ and $\tilde{H}_{h}$.}
\label{fig:lat}
\end{figure}

In order to bring the IDWM into a more tractable form, we apply the following unitary transformation,
\begin{equation*}
U_{1}=\prod_{n}\exp\Big(\frac{i\pi}{4}(\sigma^{z}_{2n}\sigma^{z}_{2n+1}-\sigma^{z}_{2n})\Big),
\end{equation*}
which is similar to the self-duality transformation in Eq.~(\ref{eq:selfdual}) but acts on every other bond and site. It maps the IDWM into $U_{1}^{\dagger}H_{0}U_{1}$, which contains only nearest-neighbor interactions~\footnotemark[\thefnnumber]. We then apply another unitary transformation composed of single-site operators,
\begin{equation*}
\begin{split}
U_{2} = & \prod_{l}\sigma^{z}_{l}\prod_{n}R_{3n+1}^{\dagger}\Big([111],\frac{2\pi}{3}\Big)R_{3n+2}\Big([111],\frac{2\pi}{3}\Big) \\
  \cdot & \prod_{k}\sigma^{z}_{6k}\sigma^{x}_{6k+1}\sigma^{x}_{6k+2}\sigma^{y}_{6k+3}\sigma^{y}_{6k+4}\sigma^{z}_{6k+5},
\end{split}
\end{equation*}
where $R_{l}([111],2\pi/3)=\exp\big(i\pi(\sigma^{x}_{l}+\sigma^{y}_{l}+\sigma^{z}_{l})/3\sqrt{3}\big)$ is the spin rotation around the $[111]$ axis by $2\pi/3$ and permutes the Pauli operators cyclically, $R_{l}^{\dagger}(\sigma^{x}_{l},\sigma^{y}_{l},\sigma^{z}_{l})R_{l}=(\sigma^{y}_{l},\sigma^{z}_{l},\sigma^{x}_{l})$. This maps the IDWM into a deformed Heisenberg chain, $\tilde{H}_{0}=U_{2}^{\dagger}U_{1}^{\dagger}H_{0}U_{1}U_{2}$,
\begin{equation}
\tilde{H}_{0} =\sum_{l\in\gamma}(\sigma_{l}^{\alpha}\sigma_{l+1}^{\alpha}+\sigma_{l}^{\beta}\sigma_{l+1}^{\beta}),
\label{eq:heis}
\end{equation}
where $\gamma$ denotes the bond index shown in Fig.~\ref{fig:lat}, which has a three-site periodicity, and $(\alpha,\beta,\gamma)$ form a right-handed frame.

The deformed Heisenberg chain in Eq.~(\ref{eq:heis}) has exactly the same form as the Kitaev-Gamma chain in a rotated basis studied in Ref.~\cite{Yang2020h}, and has a nonsymmorphic octahedral symmetry, which we explain below. It has a three-site translation symmetry $T_{3}$ ($T_{n}$ denotes the lattice translation by $n$ sites), three global spin flip symmetries $\mcal{R}(\hat{n},\pi)=\prod_{l}R_{l}(\hat{n},\pi)$ ($\hat{n}=\hat{x}$, $\hat{y}$, $\hat{z}$), and an antiunitary time-reversal symmetry $\mcal{T}$. Moreover, it is also invariant under the spin-screw translation $\mcal{R}_{T}T_{1}$ and the spin-flip reflection $\mcal{R}_{I}I_{0}$, where $\mcal{R}_{T}=\prod_{l}R_{l}([111],-2\pi/3)$ is a spin rotation, $\mcal{R}_{I}=\prod_{l}R_{l}([1\bar{1}0],\pi)$ is a spin flip, and $I_{0}$ denotes the site-$0$-centered lattice reflection. These symmetry operators generate a nonsymmorphic group $G_{0}$, which has a normal subgroup generated by the three-site translation $\langle T_{3}\rangle\cong \mbb{Z}$, and $G_{0}/\langle T_{3}\rangle\cong O_{h}$, where $O_{h}$ denotes the full octahedral group~\cite{Yang2020h}.

It has been shown that the low-energy physics of the deformed Heisenberg chain Eq.~(\ref{eq:heis}) is described by the SU(2)$_{1}$ WZW theory~\cite{Yang2020h}. Let us briefly recapitulate the arguments below. In the non-Abelian bosonization formalism, the spin operators are mapped to the following continuum fields~\cite{Affleck1986, Gogolin2004},
\begin{equation}
\mbf{S}(x)\simeq \mbf{J}(x)+\bar{\mbf{J}}(x)+(-1)^{x}\mbf{N}(x),
\label{eq:bosonization}
\end{equation}
where $\mbf{J}(x)$ and $\bar{\mbf{J}}(x)$ are the right- and the left-moving current operators, respectively, and $\mbf{N}(x)$ is the staggered component of the local spin density. The Heisenberg chain is described by the following Hamiltonian density in the continuum limit~\cite{Affleck1986, Gogolin2004},
\begin{equation}
\mcal{H}_{0}\simeq \frac{2\pi}{3}v(\mbf{J}\cdot\mbf{J}+\bar{\mbf{J}}\cdot\bar{\mbf{J}})+\lambda \mbf{J}\cdot\bar{\mbf{J}},
\label{eq:WZW}
\end{equation}
where $v$ is the velocity of low-energy excitations and $\lambda$ is a nonuniversal coupling constant. The first term is exactly the SU(2)$_{1}$ WZW model, while the second term is an SU(2) invariant perturbation, which is marginally irrelevant for $\lambda<0$. It turns out that these two terms are the only relevant or marginal operators compatible with the nonsymmorphic octahedral symmetry~\cite{Yang2020h}. For example, the current operator bilinear terms that break the SU(2) symmetry are prohibited by $\mcal{R}_{T}T_{1}$ and the global spin flip symmetries, while the bond dimerization term $\epsilon$ is prohibited by $\mcal{R}_{T}T_{1}$ and $\mcal{R}_{I}I_{0}$ symmetries. Therefore, the IDWM has an emergent SU(2)$_{1}$ conformal symmetry in the low-energy limit and is described by the SU(2)$_{1}$ WZW theory.

{\it Domain wall transition.---} A weak uniform magnetic field in the bulk introduces a nonzero string tension of the $\mbb{Z}_{2}$ gauge field described by $h\sum_{l}\tau_{l}^{z}$, where $\tau_{l}^{z}$ acts on the bulk spins. While it is an irrelevant perturbation to the bulk topological orders because of the bulk energy gap, it turns out to drive a domain wall transition. Following the derivation of the IDWM, this term is  mapped to an Ising interaction in the domain wall model~\footnotemark[\thefnnumber],
\begin{equation*}
H_{h}=-h\sum_{l}\sigma^{z}_{l}\sigma^{z}_{l+1}.
\end{equation*}
It respects the self-duality symmetry and the antiunitary symmetry, $\mcal{S}^{\dagger}H_{h}\mcal{S}=H_{h}$, $\mcal{A}^{-1}H_{h}\mcal{A}=H_{h}$.

Applying the same unitary transformations $U_{1}U_{2}$, we find
\begin{equation*}
\tilde{H}_{h}=h\sum_{l\in\gamma}\sigma_{l}^{\beta}\sigma_{l+1}^{\alpha},
\end{equation*}
where we adopt the same notations of bond and spin operator indices illustrated in Fig.~\ref{fig:lat}. While the global spin flip symmetries $\mcal{R}(\hat{n},\pi)$ ($\hat{n}=\hat{x},\hat{y},\hat{z}$) are explicitly broken, the nonsymmorphic $\mcal{R}_{T}T_{1}$ and $\mcal{R}_{I}I_{0}$ and the time-reversal symmetry $\mcal{T}$ are still preserved. The symmetry group $G_{1}$ generated by these operators satisfies $G_{1}/\langle T_{3}\rangle\cong D_{3h}$.

With the bosonization formula~(\ref{eq:bosonization}) and the operator product expansion (OPE) relations listed in the SM~\footnotemark[\thefnnumber], $H_{h}$ is mapped to the following Hamiltonian density,
\begin{equation*}
\mcal{H}_{h}\simeq \frac{4}{3}h (J_{z}\bar{J}_{y}+\bar{J}_{z}J_{y}+J_{y}\bar{J}_{x}+\bar{J}_{y}J_{x}+J_{x}\bar{J}_{z}+\bar{J}_{x}J_{z}),
\end{equation*}
where terms proportional to $\mbf{J}$ and $\bar{\mbf{J}}$ are omitted because they are conserved currents and can be eliminated by redefining the boson field $\varphi_{s}$ introduced below. This symmetric bilinear term can be simplified by an orthogonal transformation, $\mbf{M}=V\mbf{J}$ and $\bar{\mbf{M}}=V\bar{\mbf{J}}$, where $V=R(\hat{x},\arctan \sqrt{2})R(\hat{z},\pi/4)$ rotates the $[111]$ axis into the $\hat{z}$ axis, and we find
\begin{equation*}
\mcal{H}_{h}\simeq \frac{4}{3}h(2M_{z}\bar{M}_{z}-M_{x}\bar{M}_{x}-M_{y}\bar{M}_{y}).
\end{equation*}
It has an emergent U(1) symmetry of rotation around the new $\hat{z}$ axis, which is enhanced from the nonsymmorphic $D_{3h}$ symmetry of the lattice model.

\begin{figure}[tb]
\centering
\includegraphics[width=\columnwidth]{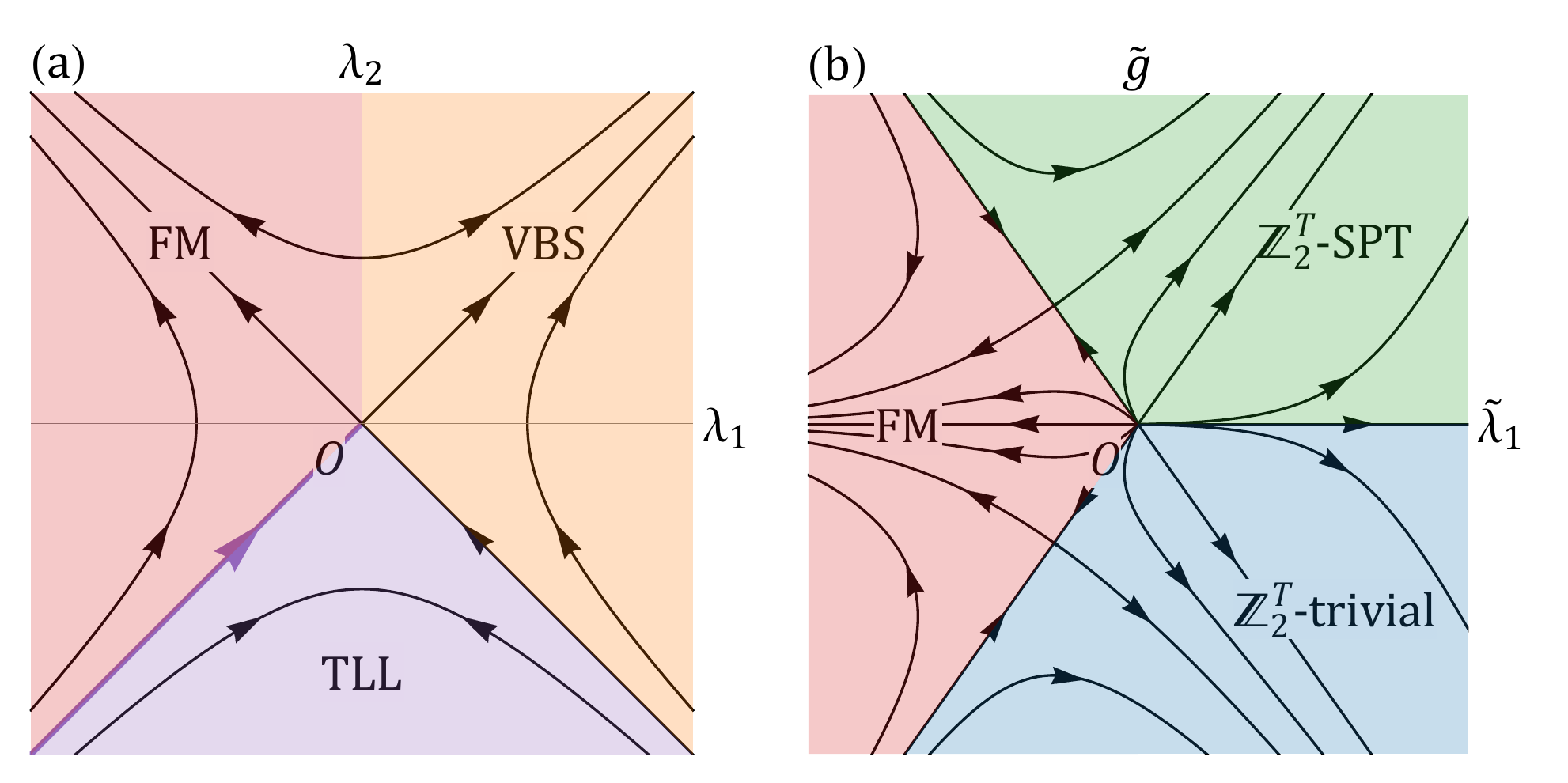}
\caption{Perturbative RG flow of (a) the domain wall model $\mcal{H}_{1}=\mcal{H}_{0}+\mcal{H}_{h}$ dictated by Eq.~(\ref{eq:rg1}) and (b) the spin chain $\mcal{H}_{2}=\mcal{H}_{1}+\mcal{H}_{g}$ given by Eq.~(\ref{eq:rg2}) in the $K_{s}=0$ limit. In (a), the region in orange is the valence-bond solid (VBS) phase of the AF Heisenberg chain~\cite{Sachdev2011}, which is not realized in the domain wall model.}
\label{fig:rg}
\end{figure}

The perturbed domain wall Hamiltonian $\mcal{H}_{1}=\mcal{H}_{0}+\mcal{H}_{h}$ is given by
\begin{equation}
\begin{split}
\mcal{H}_{1} \simeq & \frac{2\pi}{3}v(\mbf{M}\cdot\mbf{M}+\bar{\mbf{M}}\cdot\bar{\mbf{M}}) \\
+& \lambda_{1}(M_{x}\bar{M}_{x}+M_{y}\bar{M}_{y})+\lambda_{2}M_{z}\bar{M}_{z},
\end{split}
\label{eq:h1}
\end{equation}
where $\lambda_{1}=\lambda-4h/3$ and $\lambda_{2}=\lambda+8h/3$. The perturbative renormalization group (RG) equations are derived from the OPE relations in the SM~\footnotemark[\thefnnumber]~\cite{Cardy1996a},
\begin{equation}
\frac{\dd \lambda_{1}}{\dd l} = \frac{1}{2\pi}\lambda_{1}\lambda_{2}, \quad
\frac{\dd \lambda_{2}}{\dd l} = \frac{1}{2\pi}\lambda_{1}^{2}.
\label{eq:rg1}
\end{equation}

The RG flow dictated by Eq.~(\ref{eq:rg1}) is plotted in Fig.~\ref{fig:rg}~(a). The bare value of $\lambda$ is negative, thus the IDWM with $h=0$ flows along the purple line towards the SU(2)$_{1}$ fixed point at $\lambda_{1}=\lambda_{2}=0$. The domain wall perturbed by the Ising interaction $H_{h}$ deviates from this line. For $h<0$, the bare parameters satisfy $\lambda_{2}<\lambda_{1}<0$, and flow to the fixed point on the lower half of the vertical axis, which corresponds to a gapless TLL state. For $h>0$, the bare parameters satisfy $\lambda_{1}<\lambda_{2}<0$, and $\lambda_{1}\rightarrow -\infty$ while $\lambda_{2}\rightarrow+\infty$ with the RG flow, which implies a gapped phase.

The nature of the $h>0$ phase can be further clarified with the Abelian bosonization formalism, in which the Hamiltonian is mapped to the sine-Gordon model~\footnotemark[\thefnnumber],
\begin{equation*}
\mcal{H}_{1}\simeq \frac{1}{2}v \big(K_{s}\pi_{s}^{2}+K_{s}^{-1}(\pd_{x}\varphi_{s})^{2}\big)-\frac{\lambda_{1}}{4\pi^{2}}\cos\sqrt{8\pi}\varphi_{s},
\end{equation*}
where $\varphi_{s}$ and $\pi_{s}$ are the boson field and its conjugate momentum, respectively, and $K_{s}\simeq 1-\lambda_{2}/(4\pi v)$ is the bare Luttinger parameter. In the $h>0$ phase, $\lambda_{1}\rightarrow-\infty$ with the RG flow, hence the boson field $\varphi_{s}$ obtains a nonzero expectation value at the ground state $\langle\varphi_{s}\rangle=\pm\sqrt{\pi/8}$. It spontaneously breaks the time-reversal symmetry $\mcal{T}:\varphi_{s}\mapsto -\varphi_{s}$, and induces an AF order $\langle N_{z}\rangle\propto\sin\sqrt{2\pi}\langle\varphi_{s}\rangle$ of the deformed Heisenberg chain in the $M$-spin current representation. In the original domain wall model representation, this corresponds to the FM order $\langle \sigma^{z}_{l}\rangle\neq 0$, and spontaneously breaks both the antiunitary $\mbb{Z}_{2}^{T}$ symmetry and the $\mbb{Z}_{2}$ self-duality symmetry~\footnotemark[\thefnnumber]. This reflects the anomalous nature of the $\mbb{Z}_{2}$ symmetry: the domain wall must either be gapless or spontaneously break this symmetry.

\begin{figure}[tb]
\centering
\includegraphics[width=\linewidth]{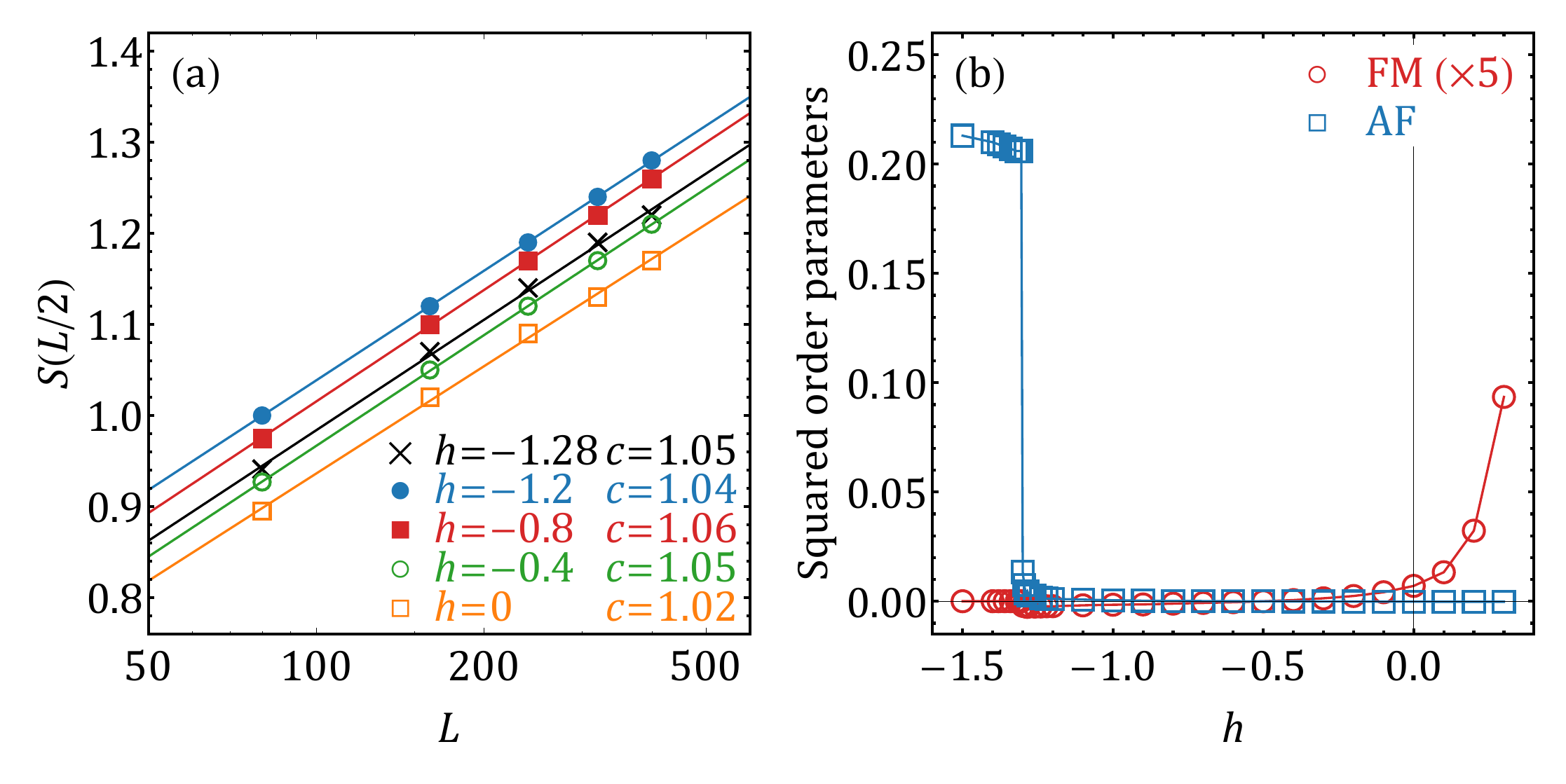}
\caption{Numerical results of the domain wall model $H_{1}$. (a) Half-chain EE $S(L/2)$ versus the chain length $L$ in the logarithmic scale in the parameter range $-1.28\leq h\leq 0$. Solid lines are the fitting to $S(L/2)=(c/6)\ln L+a$, which give an estimate of central charges listed in the legend. (b) Squared FM and AF order parameters $m_{\mr{FM}}^{2}$ and $m_{\mr{AF}}^{2}$ versus $h$ for $L=400$. All calculations are performed with the open boundary condition.}
\label{fig:hs}
\end{figure}

We perform density-matrix renormalization group (DMRG) calculations on the domain wall model $H_{1}=H_{0}+H_{h}$. The half-chain entanglement entropy (EE) $S(L/2)$ in the parameter range $-1.28\leq h\leq 0$ is plotted in Fig.~\ref{fig:hs}~(a). It obeys the formula $S(L/2)=(c/6)\ln L+a$~\cite{Calabrese2004} and gives an estimate of the central charge $c\simeq 1$ as listed in the legend, which is a hallmark of the TLL state. The squared FM and AF order parameters are estimated with the spin correlation functions, $m_{\mr{FM}}^{2}=(2/L)\sum_{r=0}^{L/2}\langle \sigma_{L/2-r/2}^{z}\sigma_{L/2+r/2}^{z}\rangle$ and $m_{\mr{AF}}^{2}=(2/L)\sum_{r=0}^{L/2}(-1)^{r}\langle \sigma_{L/2-r/2}^{z}\sigma_{L/2+r/2}^{z}\rangle$, and are plotted in Fig.~\ref{fig:hs}~(b). The TLL undergoes a continuous transition at $h\simeq 0$ to the FM order, which is consistent with the RG analysis. However, we find a transition at $h\simeq -1.3$ from the TLL to the AF order, which is not captured by the RG analysis. The abrupt change of the AF order parameter indicates that this is a first-order transition. The AF order spontaneously breaks the time-reversal, the lattice translation and the $\mbb{Z}_{2}$ self-duality symmetries, hence it is also a manifestation of the anomaly of $\mbb{Z}_{2}$ symmetry.

{\it Gapless domain wall as 1D topological transition.---} In a holographic construction~\cite{Chen2013g, Tsui2015a}, topological QCPs between SPT phases can be realized as gapless domain wall states between topological phases in one-higher dimensions. In this section, we demonstrate this holographic scenario with the gapless domain wall of $\mbb{Z}_{2}$ topological orders. We generalize the domain wall model $H_{1}$ and introduce an extra tuning parameter $g$,
\begin{equation*}
H_{2}=-\sum_{l}\big((1+g)\sigma^{z}_{l-1}\sigma^{x}_{l}\sigma^{z}_{l+1}+(1-g)\sigma^{y}_{l}+h\sigma^{z}_{l}\sigma^{z}_{l+1}\big).
\end{equation*}
The $g$-term changes sign under the self-duality transformation, hence it explicitly breaks the anomalous $\mbb{Z}_{2}$ symmetry and cannot be realized on the domain wall of $\mbb{Z}_{2}$ topological orders. Nonetheless, the antiunitary $\mbb{Z}_{2}^{T}$ symmetry $\mcal{A}=\prod_{l}\sigma_{l}^{z}\mcal{T}$ is still preserved.

Let us first highlight several exactly solvable limits. Setting $g=-1$, the model reduces to the transverse-field Ising chain, which is exactly solvable with the Jordan-Wigner transformation~\cite{Pfeuty1970}. Its ground state at $h=0$ is a direct-product state. This topologically trivial phase expands the range $-2<h<2$, and undergoes a continuous transition at $h=2$ to the FM order and another continuous transition at $h=-2$ to the AF order. Both QCPs belong to the 2D Ising universality class. On the other hand, setting $g=+1$, the model reduces to the 1D cluster Ising model, which is also exactly solvable with the Jordan-Wigner transformation~\cite{Keating2004, Jones2019}. Its ground state at $h=0$ is a $\mbb{Z}_{2}^{T}$-SPT phase, which expands the range $-2<h<2$, and undergoes 2D Ising transitions to the FM and the AF orders at $h=\pm 2$, respectively.

Applying the above unitary transformations $U_{1}U_{2}$, the $g$-term is mapped to a bond-alternating interaction,
\begin{equation*}
\tilde{H}_{g}=-g\sum_{l\in\gamma}(-1)^{l}(\sigma_{l}^{\alpha}\sigma_{l+1}^{\alpha}+\sigma_{l}^{\beta}\sigma_{l+1}^{\beta}).
\end{equation*}
This term explicitly breaks the nonsymmorphic $\mcal{R}_{T}T_{1}$ and $\mcal{R_{I}}I_{0}$ symmetries, hence allows the bond dimerization operator $\epsilon\propto \cos\sqrt{2\pi}\varphi_{s}$ in the effective theory,
\begin{equation*}
\mcal{H}_{g}\simeq \frac{g\gamma}{\pi^{2}}\cos\sqrt{2\pi}\varphi_{s},
\end{equation*}
where $\gamma\neq 0$ is a nonuniversal constant~\footnotemark[\thefnnumber]. The scaling dimension of the $\mcal{H}_{g}$ term is $1/2$ in the SU(2)$_{1}$ WZW theory, thereby it is a relevant perturbation and drives the IDWM immediately into the $\mbb{Z}_{2}^{T}$-SPT phase for $g>0$ and into the trivial phase for $g<0$. Therefore, the IDWM is a 1D topological QCP between the $\mbb{Z}_{2}^{T}$-SPT and the trivial phases.

\begin{figure}[tb]
\centering
\includegraphics[width=\columnwidth]{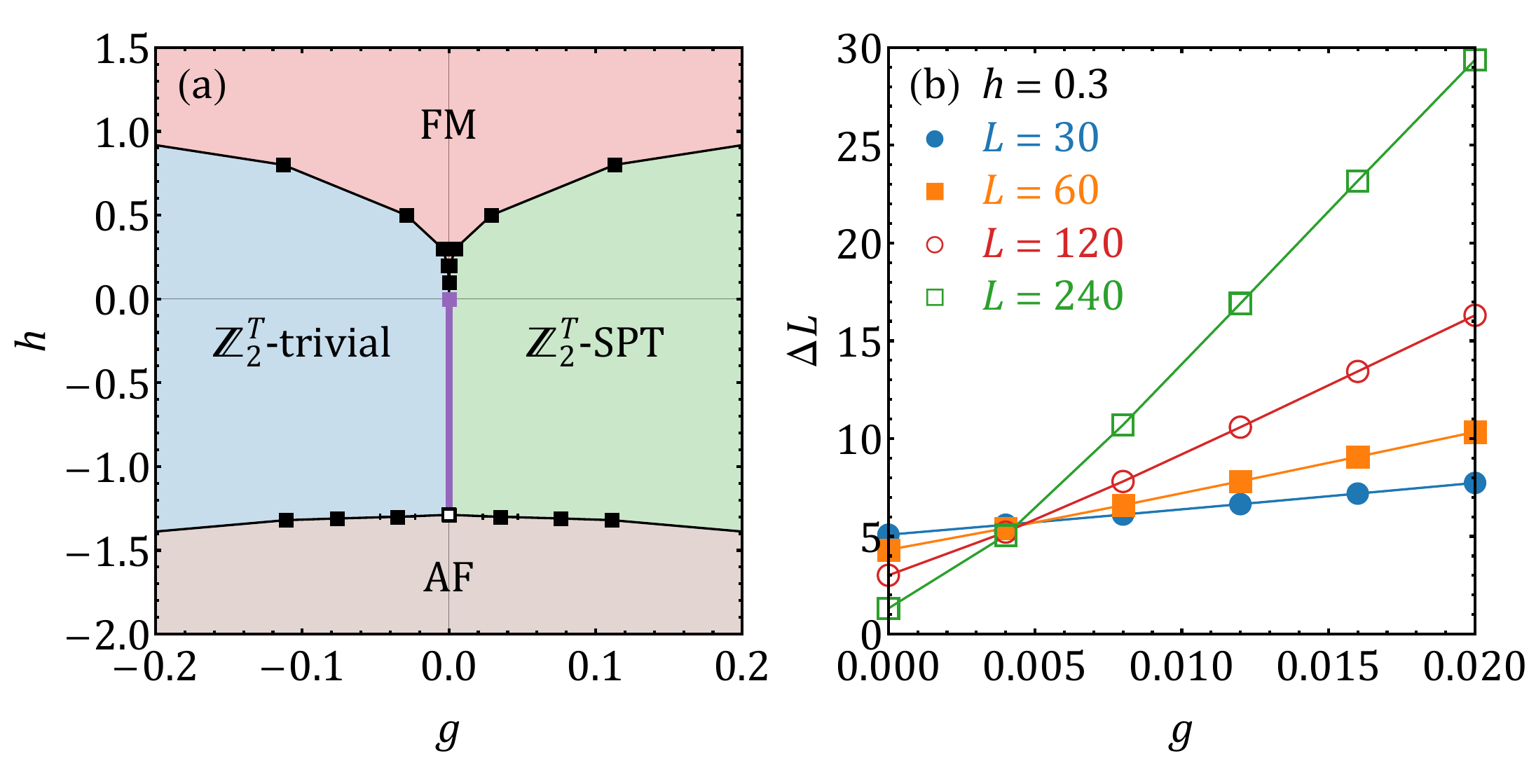}
\caption{(a) Quantum phase diagram of $H_{2}$. The purple segment on the vertical axis indicates the TLL phase of the domain wall model. Phase boundaries are extracted with the crossing of $\Delta L^{z}$ evaluated with different lattice sizes, for which an example at $h=0.3$ is shown in (b).}
\label{fig:phase}
\end{figure}

The spin chain $H_{2}$ is described by the two-frequency sine-Gordon model $\mcal{H}_{2}=\mcal{H}_{1}+\mcal{H}_{g}$. The perturbative RG equations are given by~\footnotemark[\thefnnumber]
\begin{subequations}
\label{eq:rg2}
\begin{gather}
\frd{l}{K_{s}} = -K_{s}^{2}(4\pi^{2}\tilde{\lambda}_{1}^{2}+\pi^{2}\tilde{g}^{2}), \\
\frd{l}{\tilde{\lambda}_{1}} = (2-2K_{s})\tilde{\lambda}_{1}+\frac{1}{2}\pi \tilde{g}^{2}, \\
\frd{l}{\tilde{g}} = \Big(2-\frac{1}{2}K_{s}\Big)\tilde{g}+\pi\tilde{\lambda}_{1}\tilde{g},
\end{gather}
\end{subequations}%
where the rescaled parameters $\tilde{\lambda}_{1}=\lambda_{1}/(8\pi^{2}v)$ and $\tilde{g}=g\gamma/(2\pi^{2}v)$ are introduced for convenience. For $g=0$, the RG equations are consistent with Eq.~(\ref{eq:rg1}) in the perturbative regime with $\lambda_{1,2}\ll 1$. Starting from the bare parameters $\tilde{\lambda}_{1}<0$ and $K_{s}>0$, the RG flow ends up in either the TLL phase ($K_{s}\rightarrow K_{s}^{*}\geq 1$, $\tilde{\lambda}_{1}\rightarrow0$) or the FM order ($K_{s}\rightarrow 0$, $\tilde{\lambda}_{1}\rightarrow-\infty$). The $g$-term significantly changes the RG flow. In particular, the TLL state with $K_{s}^{*}<4$ is destabilized by the relevant $g$-term, which drives $K_{s}$ to zero. Focusing on this regime and setting $K_{s}=0$ in the RG equations, we plot the RG flow of $\tilde{\lambda}_{1}$ and $\tilde{g}$ in Fig.~\ref{fig:rg}~(b), where three gapped phases are illustrated: the FM order ($\tilde{g}\rightarrow 0$, $\tilde{\lambda}_{1}\rightarrow-\infty$) spontaneously breaks the $\mbb{Z}_{2}^{T}$ symmetry, while the other two phases ($\tilde{\lambda}_{1}\rightarrow +\infty$, $\tilde{g}\rightarrow\pm\infty$) are disordered and correspond to the $\mbb{Z}_{2}^{T}$-SPT phase and trivial phase, respectively. There is a continuous transition between the FM order and either of the disordered phases, which, according to the nonperturbative analysis of the two-frequency sine-Gordon model, belongs to the 2D Ising universality class~\cite{Delfino1998, Bajnok2001}, and thus is smoothly connected with the exactly solvable QCPs of the quantum Ising chains discussed above. In addition, according to Eq.~(\ref{eq:rg2}), the TLL phase with $K_{s}^{*}>4$ should be stable against the $g$-term perturbation. However, this is not realized in the lattice model with a nonzero $g$-term according to the following numerical results. This also reflects the inadequacy of the perturbative RG analysis.

The quantum phase diagram is obtained with DMRG and plotted in Fig.~\ref{fig:phase}~(a). It is symmetric about the vertical axis, because the $g$-term flips sign under the unitary self-duality transformation. The phase boundaries are extracted with the crossing of $\Delta L^{z}$, where $\Delta$ is the energy gap, and $z=1$ is the dynamical exponent. $\Delta L^{z}$ is scale invariant at the QCP, thus its values for different lattice sizes should cross at the QCP, which is exemplified in Fig.~\ref{fig:phase}~(b). This model has four phases: the FM and the AF ordered phases, and the $\mbb{Z}_{2}^{T}$-SPT and the trivial gapped phases. The ordered phases are separated from the disordered phases by Ising critical lines. Moreover, a direct topological transition between the $\mbb{Z}_{2}^{T}$-SPT and the trivial phases takes place at the purple segment in the vertical axis, which corresponds to the gapless TLL state on the domain wall of $\mbb{Z}_{2}$ topological orders. Therefore, this model realizes the holographic scenario of topological transitions.

{\it Conclusion.---} We have theoretically investigated the IDWM of $\mbb{Z}_{2}$ topological orders and its generalization. We found the equivalence of the IDWM and the deformed Heisenberg chain with a nonsymmorphic octahedral symmetry, and thus unveiled the microscopic origin of the emergent SU(2)$_{1}$ conformal symmetry. The bulk magnetic field induces domain wall transitions from the TLL phase to the FM and the AF orders. The domain wall is either gapless or symmetry-breaking, which reflects the $\mbb{Z}_{2}$ symmetry anomaly. Moreover, the TLL state is a holographic realization of 1D topological QCP between the $\mbb{Z}_{2}^{T}$-SPT and trivial phases as the gapless domain wall of 2D topological orders. Therefore, our work provides a valuable model demonstrating the rich physics on the domain wall of topological orders.

It is desirable to generalize this work and further explore the anomalous symmetry and its consequences on the boundary and domain walls of topological phases with concrete lattice models. The interplay of the bulk topological orders with the boundary and domain wall states also deserves further study, including the generalized symmetry generated by bulk topological excitations and defects acting on the boundary and domain wall states~\cite{Chatterjee2023}, and novel boundary and domain wall critical behavior induced by the interactions of the gapless boundary and domain wall states with the bulk states at the bulk QCP.

\begin{acknowledgments}
We are grateful to helpful discussions with Zheng-Cheng Gu and Chao Xu. Part of the numerical simulations were carried out with the ITensor package \cite{Fishman2022a, Fishman2022b}. This work was supported by the National Natural Science Foundation of China (12174387 and 12304182), the Chinese Academy of Sciences (YSBR-057 and JZHKYPT-2021-08), and the Innovative Program for Quantum Science and Technology (2021ZD0302600).
\end{acknowledgments}

\bibliography{C:/Documents/ResearchNotes/BibTex/library}
\end{document}


\title{Supplemental Materials for ``Emergent Symmetry and Phase Transitions on the Domain Wall of $\mathbb{Z}_{2}$ Topological Orders''}

\author{Hong-Hao Song}
\thanks{These authors contributed equally.}
\affiliation{Kavli Institute for Theoretical Sciences and CAS Center for Excellence in Topological Quantum Computation, University of Chinese Academy of Sciences, Beijing 100190, China}

\author{Chen Peng}
\thanks{These authors contributed equally.}
\affiliation{Kavli Institute for Theoretical Sciences and CAS Center for Excellence in Topological Quantum Computation, University of Chinese Academy of Sciences, Beijing 100190, China}

\author{Rui-Zhen Huang}
\affiliation{Graduate School of China Academy of Engineering Physics, Beijing 100193, China}

\author{Long Zhang}
\email{longzhang@ucas.ac.cn}
\affiliation{Kavli Institute for Theoretical Sciences and CAS Center for Excellence in Topological Quantum Computation, University of Chinese Academy of Sciences, Beijing 100190, China}

\date{\today}

\maketitle 

\section{Unitary transformations of Ising domain wall model and its generalization}

\subsection{Ising domain wall model}

\begin{figure}[!tb]
\centering
\includegraphics[width=0.6\textwidth]{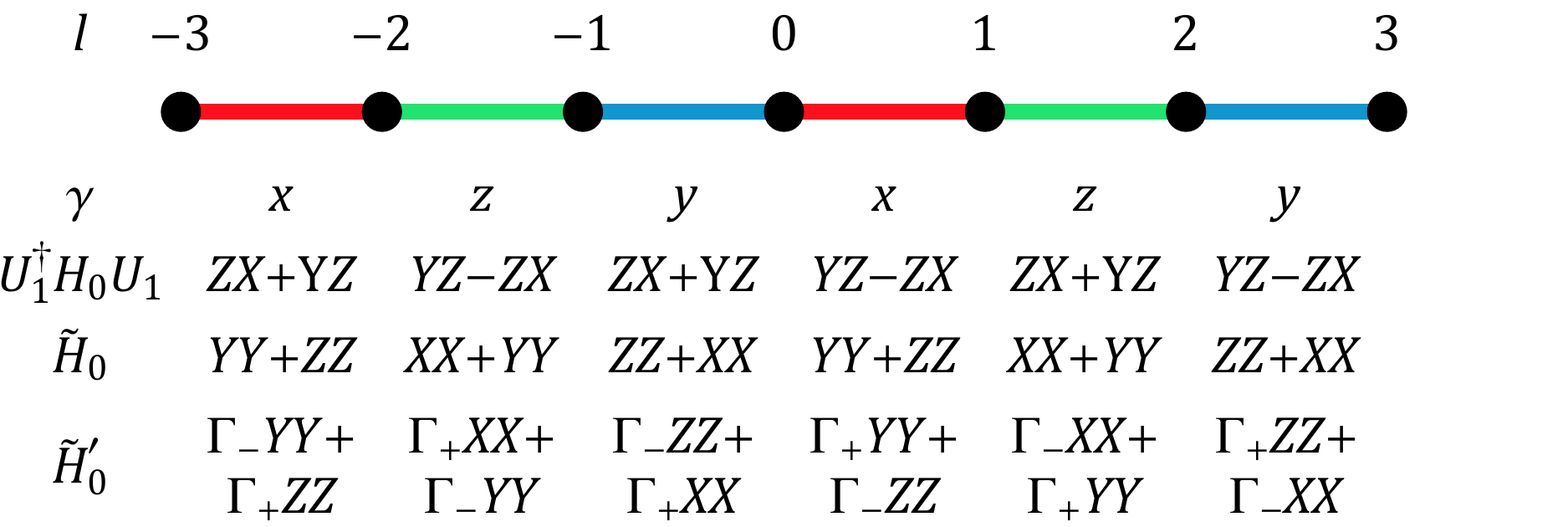}
\caption{Schematic illustration of the domain wall models in different representations.}
\label{fig:lattice}
\end{figure}

The one-dimensional (1D) domain wall of the 2D $\mbb{Z}_{2}$ topological orders with an equal gap of topological excitations is described by the Ising domain wall model (IDWM)~\cite{Bao2022},
\begin{equation}
H_{0}=-\sum_{l}\sigma^{z}_{l-1}\sigma^{x}_{l}\sigma^{z}_{l+1}-\sum_{l}\sigma^{y}_{l},
\label{eq:idw}
\end{equation}
where $\sigma^{x}_{l}$, $\sigma^{y}_{l}$ and $\sigma^{z}_{l}$ are the Pauli operators on site $l$. Applying the unitary transformation
\begin{equation}
U_{1}=\prod_{n}\exp\Big(\frac{i\pi}{4}(\sigma^{z}_{2n}\sigma^{z}_{2n+1}-\sigma^{z}_{2n})\Big),
\end{equation}
which acts on every other bond and site, the IDWM is mapped into
\begin{equation}
U_{1}^{\dagger}H_{0}U_{1}=\sum_{n}(\sigma^{z}_{2n-1}\sigma^{x}_{2n}+\sigma^{y}_{2n-1}\sigma^{z}_{2n}+\sigma^{y}_{2n}\sigma^{z}_{2n+1}-\sigma^{z}_{2n}\sigma^{x}_{2n+1}),
\end{equation}
which contains only nearest-neighbor interactions as illustrated in Fig.~\ref{fig:lattice}. We then apply another unitary transformation composed of single-site operators,
\begin{equation}
U_{2} = \prod_{l}\sigma^{z}_{l}\prod_{n}R_{3n+1}^{\dagger}\Big([111],\frac{2\pi}{3}\Big)R_{3n+2}\Big([111],\frac{2\pi}{3}\Big)
		\prod_{k}\sigma^{z}_{6k}\sigma^{x}_{6k+1}\sigma^{x}_{6k+2}\sigma^{y}_{6k+3}\sigma^{y}_{6k+4}\sigma^{z}_{6k+5},
\end{equation}
where $R_{l}([111],2\pi/3)=\exp\big(i\pi(\sigma^{x}_{l}+\sigma^{y}_{l}+\sigma^{z}_{l})/3\sqrt{3}\big)$ is the spin rotation around the $[111]$ axis by $2\pi/3$ and permutes the Pauli operators cyclically, $R_{l}^{\dagger}(\sigma^{x}_{l},\sigma^{y}_{l},\sigma^{z}_{l})R_{l}=(\sigma^{y}_{l},\sigma^{z}_{l},\sigma^{x}_{l})$. This maps the IDWM into a deformed Heisenberg chain, $\tilde{H}_{0}=U_{2}^{\dagger}U_{1}^{\dagger}H_{0}U_{1}U_{2}$,
\begin{equation}
\tilde{H}_{0} =\sum_{l\in\gamma}(\sigma_{l}^{\alpha}\sigma_{l+1}^{\alpha}+\sigma_{l}^{\beta}\sigma_{l+1}^{\beta}),
\label{eq:heis}
\end{equation}
where $\gamma$ denotes the bond index shown in Fig.~\ref{fig:lattice}, which has a three-site periodicity, and $(\alpha,\beta,\gamma)$ form a right-handed frame.

\subsection{Domain wall with unequal $\pi$-flux gaps}

The unitary transformations and symmetry analysis can also be applied to the domain wall between $\mbb{Z}_{2}$ topological orders with unequal bulk energy gaps of $\pi$-flux excitations. Following the derivation in Ref.~\cite{Bao2022}, the effective Hamiltonian of the deformed domain wall is given by
\begin{equation*}
H_{0}'=-\sum_{l}\big(1+\eta_{l}\delta\big)(\sigma^{z}_{l-1}\sigma^{x}_{l}\sigma^{z}_{l+1}+\sigma^{y}_{l}),
\end{equation*}
where $\eta_{l}=(-1)^{l}$. Here, $\Gamma_{\pm}=1\pm\delta$ denote the $\pi$-flux excitation gaps of the toric code and the double semion models, respectively. Applying the same unitary transformations $U_{1}U_{2}$ yields
\begin{equation*}
\tilde{H}_{0}'=\sum_{l\in\gamma}\big((1+\eta_{l}\delta)\sigma_{l}^{\alpha}\sigma_{l+1}^{\alpha}+(1-\eta_{l}\delta)\sigma_{l}^{\beta}\sigma_{l+1}^{\beta}\big),
\end{equation*}
where we adopt the same notations of bond and spin operator indices as in Eq.~(\ref{eq:heis}), which is shown in Fig.~\ref{fig:lattice}. $\tilde{H}_{0}'$ is exactly the Gamma chain with a Dzyaloshinskii-Moriya interaction in the rotated basis studied in Ref.~\cite{Yang2025}. Although the $\mcal{R}_{T}T_{1}$ symmetry is explicitly broken due to the unequal bulk gaps, the $(\mcal{R}_{T}T_{1})^{2}$ symmetry and other symmetries of the IDWM discussed in the main text are still preserved and generate a nonsymmorphic octahedral symmetry $G_{0}'$, which satisfies $G_{0}'/\langle T_{6}\rangle\cong O_{h}$. Similar symmetry analysis shows that all relevant and marginal interactions except those in Eq.~(5) of the main text are prohibited, hence the domain wall model with unequal bulk gaps also has an emergent SU(2)$_{1}$ conformal symmetry in the low-energy limit.

\section{Bosonization and operator product expansion relations}

The effective theory of spin chains with only nearest-neighbor interactions can be obtained with the following bosonization formula~\cite{Affleck1986, Gogolin2004},
\begin{equation}
\mbf{S}(x)\simeq \mbf{J}(x)+\bar{\mbf{J}}(x)+(-1)^{x}\mbf{N}(x),
\end{equation}
where $\mbf{J}(x)$ and $\bar{\mbf{J}}(x)$ are the right- and the left-moving spin currents, respectively, and $\mbf{N}(x)$ is the staggered component of the local spin density. These operators are defined with the right- and the left-moving spin-$1/2$ fermion operators $\psi_{\alpha}$ and $\bar{\psi}_{\alpha}$ ($\alpha=\uparrow,\downarrow$),
\begin{equation}
\mbf{J}(x)=\frac{1}{2}\psi_{\alpha}^{\dagger}(x)\mbf{\sigma}_{\alpha\beta}\psi_{\beta}(x),\quad
\bar{\mbf{J}}(x)=\frac{1}{2}\bar{\psi}_{\alpha}^{\dagger}(x)\mbf{\sigma}_{\alpha\beta}\bar{\psi}_{\beta}(x),\quad
\mbf{N}(x)=\frac{1}{2}\psi_{\alpha}^{\dagger}(x)\mbf{\sigma}_{\alpha\beta}\bar{\psi}_{\beta}(x)+\mr{H.c.},
\end{equation}
where $\mbf{\sigma}$ denotes the Pauli matrices. In addition, the bond dimerization operator $\epsilon(x)$ is defined by
\begin{equation}
\epsilon(x)=\frac{i}{2}\bar{\psi}_{\alpha}^{\dagger}(x)\psi_{\alpha}(x)+\mr{H.c.}
\end{equation}
These operators satisfy the following operator product expansion (OPE) relations,
\begin{gather}
J_{a}(z)J_{b}(w)\sim \frac{\delta_{ab}}{8\pi^{2}(z-w)^{2}}+\frac{i\epsilon_{abc}J_{c}(w)}{2\pi(z-w)}, \\
\bar{J}_{a}(\bar{z})\bar{J}_{b}(\bar{w})\sim \frac{\delta_{ab}}{8\pi^{2}(\bar{z}-\bar{w})^{2}}+\frac{i\epsilon_{abc}\bar{J}_{c}(\bar{w})}{2\pi(\bar{z}-\bar{w})}, \\
J_{a}(z)N_{b}(w,\bar{w})\sim \frac{i\epsilon_{abc}N_{c}(w,\bar{w})+i\delta_{ab}\epsilon(w,\bar{w})}{4\pi(z-w)}, \\
\bar{J}_{a}(\bar{z})N_{b}(w,\bar{w})\sim \frac{i\epsilon_{abc}N_{c}(w,\bar{w})-i\delta_{ab}\epsilon(w,\bar{w})}{4\pi(\bar{z}-\bar{w})}, \\
J_{a}(z)\epsilon(w,\bar{w})\sim \frac{-iN_{a}(w,\bar{w})}{4\pi(z-w)}, \\
\bar{J}_{a}(\bar{z})\epsilon(w,\bar{w})\sim \frac{iN_{a}(w,\bar{w})}{4\pi(\bar{z}-\bar{w})}, \\
N_{a}(z,\bar{z})N_{b}(w,\bar{w})\sim \frac{\gamma^{2}\delta_{ab}}{2\pi^{2}|z-w|}+\frac{i\gamma^{2}\epsilon_{abc}}{\pi}\Big(J_{c}(w)\frac{z-w}{|z-w|}+\bar{J}_{c}(\bar{w})\frac{\bar{z}-\bar{w}}{|z-w|}\Big), \\
N_{a}(z,\bar{z})\epsilon(w,\bar{w})\sim \frac{i\gamma^{2}}{\pi}\Big(J_{a}(w)\frac{z-w}{|z-w|}-\bar{J}_{a}(\bar{w})\frac{\bar{z}-\bar{w}}{|z-w|}\Big), \\
\epsilon(z,\bar{z})\epsilon(w,\bar{w})\sim \frac{\gamma^{2}}{2\pi^{2}|z-w|},
\end{gather}
where
\begin{equation}
z=v\tau-ix,\quad \bar{z}=v\tau+ix
\label{eq:zzbar}
\end{equation}
are the complex coordinates, and $v$ is the velocity of low-energy excitations. These OPE relations are used to derive the renormalization group (RG) equations of the domain wall model in the main text.

The Abelian bosonization formalism is more convenient in the absence of SU(2) symmetry. The fermion operators are represented by the boson fields $\phi_{\alpha}(x)$ and $\bar{\phi}_{\alpha}(x)$ as follows~\cite{Senechal2004},
\begin{gather}
\psi_{\alpha}(x)=\frac{\eta_{\alpha}}{\sqrt{2\pi}}e^{-i\sqrt{4\pi}\phi_{\alpha}(x)}, \quad
\psi_{\alpha}^{\dagger}(x)=\frac{\eta_{\alpha}}{\sqrt{2\pi}}e^{i\sqrt{4\pi}\phi_{\alpha}(x)}, \\
\bar{\psi}_{\alpha}(x)=\frac{\bar{\eta}_{\alpha}}{\sqrt{2\pi}}e^{i\sqrt{4\pi}\bar{\phi}_{\alpha}(x)}, \quad
\bar{\psi}_{\alpha}^{\dagger}(x)=\frac{\bar{\eta}_{\alpha}}{\sqrt{2\pi}}e^{-i\sqrt{4\pi}\bar{\phi}_{\alpha}(x)}.
\end{gather}
Here, $\eta_{\alpha}$ and $\bar{\eta}_{\alpha}$ are the Klein factors, which satisfy
\begin{equation}
\{\eta_{\alpha},\eta_{\beta}\}=2\delta_{\alpha\beta},\quad \{\bar{\eta}_{\alpha},\bar{\eta}_{\beta}\}=2\delta_{\alpha\beta}, \quad
\{\eta_{\alpha},\bar{\eta}_{\beta}\}=0,
\end{equation}
and are introduced to guarantee the anticommutation of fermion operators of different species. We adopt the following sign convention in the calculations,
\begin{equation}
\eta_{\uparrow}\eta_{\downarrow}\bar{\eta}_{\uparrow}\bar{\eta}_{\downarrow}=1.
\end{equation}
The chiral and antichiral boson fields satisfy
\begin{equation}
\langle \phi_{\alpha}(z)\phi_{\beta}(w)\rangle = -\frac{\delta_{\alpha\beta}}{4\pi}\ln(z-w), \quad
\langle \bar{\phi}_{\alpha}(\bar{z})\bar{\phi}_{\beta}(\bar{w})\rangle = -\frac{\delta_{\alpha\beta}}{4\pi}\ln(\bar{z}-\bar{w}).
\end{equation}
The following boson fields and dual fields are introduced for convenience,
\begin{gather}
\phi_{s}=\frac{1}{\sqrt{2}}(\phi_{\uparrow}-\phi_{\downarrow}), \quad
\bar{\phi}_{s}=\frac{1}{\sqrt{2}}(\bar{\phi}_{\uparrow}-\bar{\phi}_{\downarrow}), \\
\varphi_{\uparrow}=\phi_{\uparrow}+\bar{\phi}_{\uparrow}=\frac{1}{\sqrt{2}}(\varphi_{c}+\varphi_{s}), \quad
\varphi_{\downarrow}=\phi_{\downarrow}+\bar{\phi}_{\downarrow}=\frac{1}{\sqrt{2}}(\varphi_{c}-\varphi_{s}), \\
\theta_{\uparrow}=\phi_{\uparrow}-\bar{\phi}_{\uparrow}=\frac{1}{\sqrt{2}}(\theta_{c}+\theta_{s}), \quad
\theta_{\downarrow}=\phi_{\downarrow}-\bar{\phi}_{\downarrow}=\frac{1}{\sqrt{2}}(\theta_{c}-\theta_{s}).
\end{gather}
In the spin chain, the charge degrees of freedom are gapped, thus $\varphi_{c}$ takes a nonzero expectation value at the ground state $\gamma=\langle\cos\sqrt{2\pi}\varphi_{c}\rangle$. The spin current and bond operators are given by
\begin{gather}
J_{x}=\frac{i\eta_{\uparrow}\eta_{\downarrow}}{2\pi}\sin\sqrt{8\pi}\phi_{s}, \quad
J_{y}=-\frac{i\eta_{\uparrow}\eta_{\downarrow}}{2\pi}\cos\sqrt{8\pi}\phi_{s}, \quad
J_{z}=\frac{i}{\sqrt{2\pi}}\pd_{z}\phi_{s}, \\
\bar{J}_{x}=-\frac{i\bar{\eta}_{\uparrow}\bar{\eta}_{\downarrow}}{2\pi}\sin\sqrt{8\pi}\bar{\phi}_{s}, \quad
\bar{J}_{y}=-\frac{i\bar{\eta}_{\uparrow}\bar{\eta}_{\downarrow}}{2\pi}\cos\sqrt{8\pi}\bar{\phi}_{s}, \quad
\bar{J}_{z}=-\frac{i}{\sqrt{2\pi}}\pd_{\bar{z}}\bar{\phi}_{s}, \\
N_{x}=\frac{i\eta_{\uparrow}\bar{\eta}_{\downarrow}\gamma}{\pi}\sin\sqrt{2\pi}\theta_{s}, \quad
N_{y}=-\frac{i\eta_{\uparrow}\bar{\eta}_{\downarrow}\gamma}{\pi}\cos\sqrt{2\pi}\theta_{s}, \quad
N_{z}=\frac{i\eta_{\uparrow}\bar{\eta}_{\uparrow}\gamma}{\pi}\sin\sqrt{2\pi}\varphi_{s}, \quad
\epsilon=-\frac{i\eta_{\uparrow}\bar{\eta}_{\uparrow}\gamma}{\pi}\cos\sqrt{2\pi}\varphi_{s}.
\end{gather}

\subsection{Symmetry transformations}

\subsubsection{Time reversal}

Under the antiunitary time-reversal transformation $\mcal{T}$, the right- and the left-moving fermion operators exchange with each other,
\begin{gather}
\mcal{T}: 	\psi_{\uparrow}(z)\mapsto \bar{\psi}_{\downarrow}(\bar{z}), \quad
			\psi_{\downarrow}(z)\mapsto -\bar{\psi}_{\uparrow}(\bar{z}), \\
\mcal{T}:	\bar{\psi}_{\uparrow}(\bar{z})\mapsto \psi_{\downarrow}(z), \quad
			\bar{\psi}_{\downarrow}(\bar{z})\mapsto -\psi_{\uparrow}(z).
\end{gather}
The Klein factors transform by
\begin{equation}
\mcal{T}:	\eta_{\uparrow}\mapsto \bar{\eta}_{\downarrow}, \quad
			\eta_{\downarrow}\mapsto -\bar{\eta}_{\uparrow}, \quad
			\bar{\eta}_{\uparrow}\mapsto \eta_{\downarrow}, \quad
			\bar{\eta}_{\downarrow}\mapsto -\eta_{\uparrow},
\end{equation}
and we find
\begin{equation}
\mcal{T}: \eta_{\uparrow}\eta_{\downarrow}\bar{\eta}_{\uparrow}\bar{\eta}_{\downarrow}\mapsto \bar{\eta}_{\downarrow}\bar{\eta}_{\uparrow}\eta_{\downarrow}\eta_{\uparrow}=\eta_{\uparrow}\eta_{\downarrow}\bar{\eta}_{\uparrow}\bar{\eta}_{\downarrow},
\end{equation}
thus the sign convention is not changed under the time-reversal transformation. The spin current and bond operators transform as follows,
\begin{equation}
\mcal{T}:	\mbf{J}(z)\mapsto -\bar{\mbf{J}}(\bar{z}), \quad
			\bar{\mbf{J}}(\bar{z})\mapsto -\mbf{J}(z), \quad
			\mbf{N}(z,\bar{z})\mapsto -\mbf{N}(z,\bar{z}), \quad
			\epsilon(z,\bar{z})\mapsto \epsilon(z,\bar{z}).
\end{equation}
The time-reversal transformations of the boson field and the dual field are given by
\begin{equation}
\mcal{T}:	\varphi_{s}(z,\bar{z})\mapsto -\varphi_{s}(z,\bar{z}), \quad
			\theta_{s}(z,\bar{z})\mapsto \theta_{s}(z,\bar{z}).
\end{equation}

\subsubsection{Lattice translation}

The lattice translation transformations of the right- and the left-moving fermion operators are given by
\begin{equation}
T_{1}:	\psi_{\alpha}(x)\mapsto i\psi_{\alpha}(x+1), \quad
		\bar{\psi}_{\alpha}(x)\mapsto -i\bar{\psi}_{\alpha}(x+1),
\end{equation}
which can be obtained by requiring
\begin{gather}
T_{1}:	\eta_{\alpha}\mapsto \eta_{\alpha}, \quad
		\bar{\eta}_{\alpha}\mapsto \bar{\eta}_{\alpha}, \\
T_{1}: 	\sqrt{4\pi}\phi_{\alpha}(x)\mapsto \sqrt{4\pi}\phi_{\alpha}(x+1)-\pi/2, \quad
		\sqrt{4\pi}\bar{\phi}_{\alpha}(x)\mapsto \sqrt{4\pi}\bar{\phi}_{\alpha}(x+1)+\pi/2.
\end{gather}
Moreover, we find
\begin{gather}
T_{1}:	\mbf{J}(x)\mapsto \mbf{J}(x+1),\quad \bar{\mbf{J}}(x)\mapsto \bar{\mbf{J}}(x+1), \quad
		\mbf{N}(x)\mapsto -\mbf{N}(x+1), \quad \epsilon(x)\mapsto -\epsilon(x+1). \\
T_{1}:	\sqrt{2\pi}\varphi_{s}(x)\mapsto \sqrt{2\pi}\varphi_{s}(x+1)+\pi, \quad
		\sqrt{2\pi}\theta_{s}(x)\mapsto \sqrt{2\pi}\theta_{s}(x+1)+\pi.
\end{gather}

\subsubsection{Uniaxial spin rotation}

Under the spin rotation around the $z$ axis $R_{z}(\omega)$, the fermion operators obtain extra phase factors,
\begin{equation}
R_{z}(\omega):~\psi_{\alpha}(z)\mapsto e^{-i\alpha\omega/2}\psi_{\alpha}(z),\quad
\bar{\psi}_{\alpha}(\bar{z})\mapsto e^{-i\alpha\omega/2}\bar{\psi}_{\alpha}(\bar{z}),
\end{equation}
which leads to
\begin{gather}
R_{z}(\omega):~\eta_{\alpha}\mapsto \eta_{\alpha},\quad \bar{\eta}_{\alpha}\mapsto\bar{\eta}_{\alpha}, \\
R_{z}(\omega):~\sqrt{4\pi}\phi_{\alpha}(z)\mapsto \sqrt{4\pi}\phi_{\alpha}(z)-\alpha\omega/2,\quad
\sqrt{4\pi}\bar{\phi}_{\alpha}(\bar{z})\mapsto \sqrt{4\pi}\bar{\phi}_{\alpha}(\bar{z})+\alpha\omega/2.
\end{gather}
Hence we find
\begin{gather}
R_{z}(\omega):~\sqrt{8\pi}\phi_{s}(z)\mapsto\sqrt{8\pi}\phi_{s}(z)+\omega,\quad
\sqrt{8\pi}\bar{\phi}_{s}(\bar{z})\mapsto\sqrt{8\pi}\bar{\phi}_{s}(\bar{z})-\omega, \\
R_{z}(\omega):~\sqrt{2\pi}\varphi_{s}(z,\bar{z})\mapsto \sqrt{2\pi}\varphi_{s}(z,\bar{z}),\quad
\sqrt{2\pi}\theta_{s}(z,\bar{z})\mapsto \sqrt{2\pi}\theta_{s}(z,\bar{z})+\omega,
\end{gather}
and
\begin{gather}
R_{z}(\omega):~
\begin{pmatrix}
J_{x}(z) \\
J_{y}(z)
\end{pmatrix}
\mapsto
\begin{pmatrix}
\cos\omega & -\sin\omega \\
\sin\omega & \cos\omega
\end{pmatrix}
\begin{pmatrix}
J_{x}(z) \\
J_{y}(z)
\end{pmatrix}, \quad
\begin{pmatrix}
\bar{J}_{x}(\bar{z}) \\
\bar{J}_{y}(\bar{z})
\end{pmatrix}
\mapsto
\begin{pmatrix}
\cos\omega & -\sin\omega \\
\sin\omega & \cos\omega
\end{pmatrix}
\begin{pmatrix}
\bar{J}_{x}(\bar{z}) \\
\bar{J}_{y}(\bar{z})
\end{pmatrix}, \\
R_{z}(\omega):~J_{z}(z)\mapsto J_{z}(z),\quad \bar{J}_{z}(\bar{z})\mapsto \bar{J}_{z}(\bar{z}), \\
R_{z}(\omega):~
\begin{pmatrix}
N_{x}(z,\bar{z}) \\
N_{y}(z,\bar{z})
\end{pmatrix}
\mapsto
\begin{pmatrix}
\cos\omega & -\sin\omega \\
\sin\omega & \cos\omega
\end{pmatrix}
\begin{pmatrix}
N_{x}(z,\bar{z}) \\
N_{y}(z,\bar{z})
\end{pmatrix}, \\
R_{z}(\omega):~N_{z}(z,\bar{z})\mapsto N_{z}(z,\bar{z}),\quad \epsilon(z,\bar{z})\mapsto \epsilon(z,\bar{z}).
\end{gather}

\section{Mapping of ferromagnetic order parameter}
\label{sec:mapping}

\begin{table}[!tb]
\centering
\caption{The FM order in the domain wall model maps to the six-site periodic noncollinear spin order in the $\tilde{H}_{1}$ ($J$-spin current) representation. The latter has a nonzero overlap with the AF order in the $[111]$ direction, which maps to the AF Ising order in $\hat{z}$ axis in the $M$-spin current representation via the orthogonal transformation.}
\begin{ruledtabular}
\begin{tabular}{ccccccc}
$l$									& 0		& 1		& 2		& 3		& 4		& 5		\\
\hline
Domain wall	model					& $Z$	& $Z$	& $Z$	& $Z$	& $Z$	& $Z$	\\
$\tilde{H}_{1}$ ($J$-spin current)	& $Z$	& $-Y$	& $X$	& $-Z$	& $Y$	& $-X$	\\
$M$-spin current					& $Z$	& $-Z$	& $Z$	& $-Z$	& $Z$	& $-Z$
\end{tabular}
\end{ruledtabular}
\label{tab:order}
\end{table}

Let us examine the nature of the gapped phase with $h>0$ in the perturbed domain wall model. In this phase, $\lambda_{1}\rightarrow -\infty$ with the RG flow, thus the boson field $\varphi_{s}$ obtains a nonzero expectation value at the ground state, $\lr{\varphi_{s}}=\pm\sqrt{\pi/8}$. It leads to an Ising antiferromagnetic (AF) order $\lr{N_{z}}\propto \lr{\sin\sqrt{2\pi}\varphi_{s}}\neq 0$ in the $M$-spin current representation and spontaneously breaks the time reversal and the lattice translation symmetries. On the other hand, the ferromagnetic (FM) order parameter $\lr{\sigma_{l}^{z}}$ in the domain wall model is mapped into a six-site periodic noncollinear spin order in the $J$-spin current representation as listed in Table~\ref{tab:order}. This noncollinear order has a nonzero overlap with the AF order in $[111]$ direction, which exactly maps to the Ising AF order in $\hat{z}$ axis in the $M$-spin current representation via the orthogonal transformation, $\mbf{M}=V\mbf{J}$ and $\bar{\mbf{M}}=V\bar{\mbf{J}}$, where the orthogonal matrix $V=R(\hat{x},\arctan \sqrt{2})R(\hat{z},\pi/4)$ rotates the $[111]$ axis into the $\hat{z}$ axis. Therefore, the gapped phase with $h>0$ corresponds to the FM order in the perturbed domain wall model.

\section{RG equations of the two-frequency sine-Gordon model}

The generalized spin chain is described by the two-frequency sine-Gordon model,
\begin{equation}
\mcal{H}_{2}= \frac{1}{2}v\big(K_{s}\pi_{s}^{2}+K_{s}^{-1}(\pd_{x}\varphi_{s})^{2}\big)-\frac{\lambda_{1}}{4\pi^{2}}\cos\sqrt{8\pi}\varphi_{s}+\frac{g\gamma}{\pi}i\eta_{\uparrow}\eta_{\downarrow}\cos\sqrt{2\pi}\varphi_{s},
\end{equation}
where
\begin{equation}
K_{s}=\sqrt{\frac{1-\lambda_{2}/(4\pi v)}{1+\lambda_{2}/(4\pi v)}}
\end{equation}
is the bare Luttinger parameter. Performing the Legendre transformation with
\begin{equation}
\pd_{t}\varphi_{s}=\frac{\pd \mcal{H}_{2}}{\pd \pi_{s}}=vK_{s}\pi_{s},
\end{equation}
we find the following real-time Lagrangian density
\begin{equation}
\mcal{L}=\frac{v}{2K_{s}}\big(v^{-2}(\pd_{t}\varphi_{s})^{2}-(\pd_{x}\varphi_{s})^{2}\big)+\frac{\lambda_{1}}{4\pi^{2}}\cos\sqrt{8\pi}\varphi_{s}-\frac{g\gamma}{\pi}i\eta_{\uparrow}\eta_{\downarrow}\cos\sqrt{2\pi}\varphi_{s}.
\end{equation}
In the Euclidean spacetime with the imaginary time $\tau=it$, the Lagrangian density is given by
\begin{equation}
\mcal{L}_{2}=\frac{v}{2K_{s}}\big(v^{-2}(\pd_{\tau}\varphi_{s})^{2}+(\pd_{x}\varphi_{s})^{2}\big)-\frac{\lambda_{1}}{4\pi^{2}}\cos\sqrt{8\pi}\varphi_{s}+\frac{g\gamma}{\pi}i\eta_{\uparrow}\eta_{\downarrow}\cos\sqrt{2\pi}\varphi_{s}.
\end{equation}
In the complex coordinates introduced in Eq.~(\ref{eq:zzbar}) with
\begin{equation}
\pd_{z}=\frac{1}{2}(v^{-1}\pd_{\tau}+i\pd_{x}), \quad
\pd_{\bar{z}}=\frac{1}{2}(v^{-1}\pd_{\tau}-i\pd_{x}),
\end{equation}
the action is given by
\begin{equation}
S=\int\dd z\dd\bar{z}\, \big(K_{s}^{-1}\pd_{z}\varphi_{s}\pd_{\bar{z}}\varphi_{s}-\tilde{\lambda}_{1}\cos\sqrt{8\pi}\varphi_{s}+\tilde{g}\cos\sqrt{2\pi}\varphi_{s}\big),
\end{equation}
where we have introduced the rescaled parameters $\tilde{\lambda}_{1}=\lambda_{1}/(8\pi^{2} v)$ and $\tilde{g}=g\gamma/(2\pi^{2}v)$, and take the sign convention $i\eta_{\uparrow}\eta_{\downarrow}=1$ for convenience. The correlation function at the free-boson fixed point is
\begin{equation}
\langle \varphi_{s}(z,\bar{z})\varphi_{s}(w,\bar{w})\rangle=-\frac{K_{s}}{4\pi}\ln|z-w|^{2}.
\end{equation}
The perturbative RG equations can be derived based on the following OPE relations,
\begin{gather}
\begin{split}
& :\cos\sqrt{8\pi}\varphi_{s}(z,\bar{z})::\cos\sqrt{8\pi}\varphi_{s}(w,\bar{w}):
\sim \frac{1}{2|z-w|^{4K_{s}}} - \frac{4\pi}{|z-w|^{4K_{s}-2}}:\pd_{z}\varphi_{s}\pd_{\bar{z}}\varphi_{s}: \\
& - \frac{2\pi (z-w)^2}{|z-w|^{4K_{s}}}:\pd_{z}\varphi_{s}\pd_{z}\varphi_{s}:
  - \frac{2\pi (\bar{z}-\bar{w})^2}{|z-w|^{4K_{s}}}  :\pd_{\bar{z}}\varphi_{s}\pd_{\bar{z}}\varphi_{s}:
  + \frac{1}{2}|z-w|^{4K_{s}}:\cos\sqrt{32\pi}\varphi_{s}:,
\end{split}
\\
\begin{split}
& :\cos\sqrt{2\pi}\varphi_{s}(z,\bar{z})::\cos\sqrt{2\pi}\varphi_{s}(w,\bar{w}):
\sim \frac{1}{2|z-w|^{K_{s}}} - \frac{\pi}{|z-w|^{K_{s}-2}}:\pd_{z}\varphi_{s}\pd_{\bar{z}}\varphi_{s}: \\
& - \frac{\pi (z-w)^2}{2|z-w|^{K_{s}}}:\pd_{z}\varphi_{s}\pd_{z}\varphi_{s}:
  - \frac{\pi (\bar{z}-\bar{w})^2}{2|z-w|^{K_{s}}}:\pd_{\bar{z}}\varphi_{s}\pd_{\bar{z}}\varphi_{s}:
  + \frac{1}{2}|z-w|^{K_{s}}:\cos\sqrt{8\pi}\varphi_{s}:,
\end{split}
\\
\begin{split}
& :\cos\sqrt{8\pi}\varphi_{s}(z,\bar{z})::\cos\sqrt{2\pi}\varphi_{s}(w,\bar{w}):
\sim \frac{1}{2|z-w|^{2K_{s}}}:\cos\sqrt{2\pi}\varphi_{s}: + \frac{1}{2}|z-w|^{2K_{s}}:\cos \sqrt{18\pi}\varphi_{s}:,
\\
\end{split}
\end{gather}
and we find
\begin{gather}
\frd{l}{K_{s}} = -K_{s}^{2}(4\pi^{2}\tilde{\lambda}_{1}^{2}+\pi^{2}\tilde{g}^{2}), \\
\frd{l}{\tilde{\lambda}_{1}} = (2-2K_{s})\tilde{\lambda}_{1}+\frac{1}{2}\pi \tilde{g}^{2}, \\
\frd{l}{\tilde{g}} = \Big(2-\frac{1}{2}K_{s}\Big)\tilde{g}+\pi\tilde{\lambda}_{1}\tilde{g}.
\end{gather}

\section{More numerical results}

\subsection{Mutual information of IDWM}

In a 1D quantum critical state described by conformal field theory (CFT), the entanglement entropy (EE) of an interval embedded in a chain with periodic boundary condition (PBC) is given by~\cite{Calabrese2004},
\begin{equation}
S(r,L)=\frac{c}{3}\ln\Big(\frac{L}{\pi}\sin\frac{\pi r}{L}\Big),
\end{equation}
where $r$ and $L$ are the lengths of the interval and the chain, respectively, and $c$ is the central charge of the CFT. This formula has been commonly adopted to extract the central charge of a quantum critical state. However, it does not distinguish the Tomonaga-Luttinger liquid (TLL) states with different Luttinger parameters, which have the same central charge $c=1$ but different operator contents.

\begin{figure}[!tb]
\centering
\includegraphics[width=0.5\textwidth]{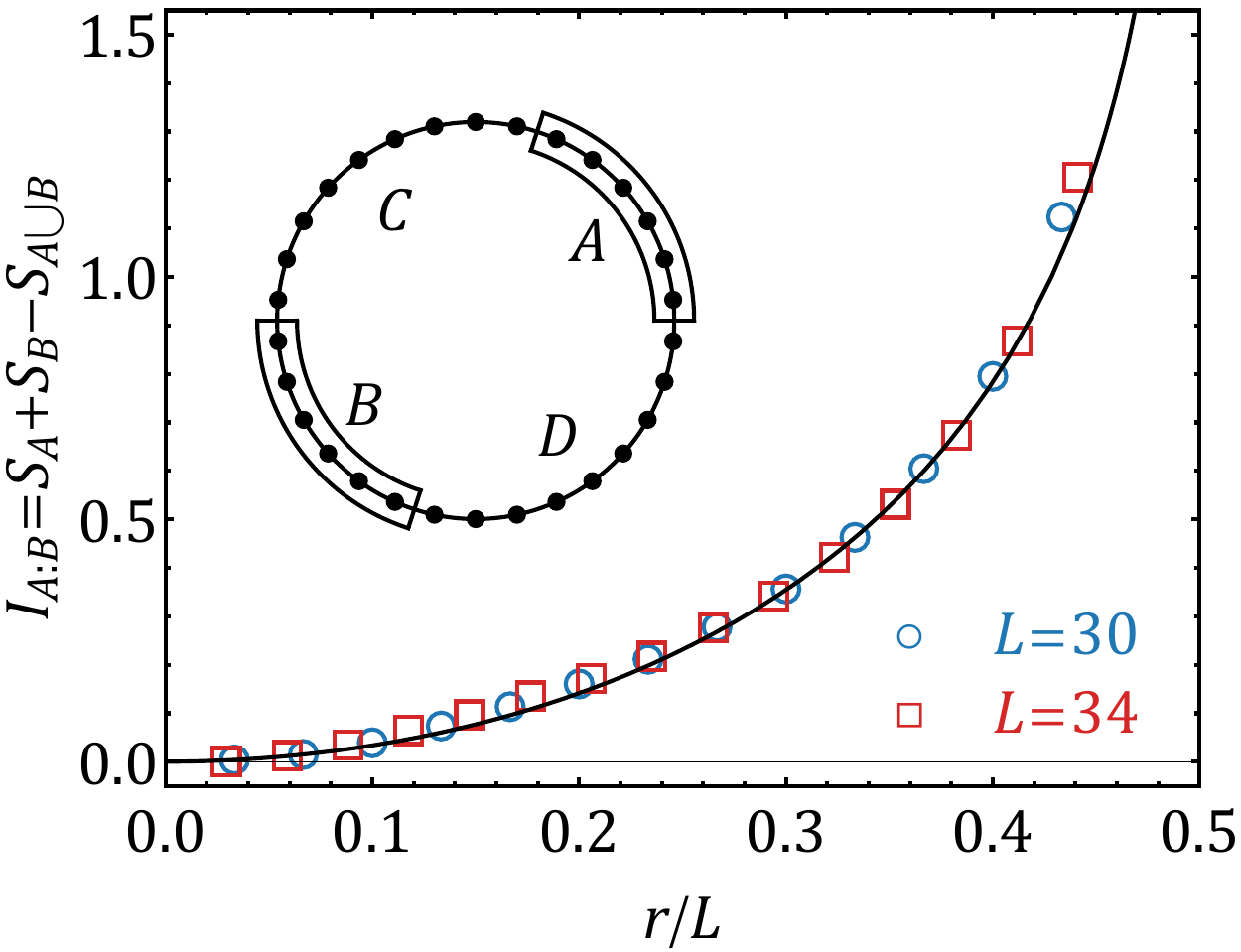}
\caption{Mutual information $I_{A:B}$ of the IDWM. The solid curve is the theoretical result of the SU(2)$_{1}$ WZW theory~\cite{Furukawa2009, Calabrese2009c}, which matches perfectly with the results of the IDWM, thus providing further support to the SU(2)$_{1}$ WZW effective theory. Inset: The geometry adopted to calculate the mutual information $I_{A:B}$.}
\label{fig:mutual}
\end{figure}

The mutual information of two disjoint intervals is sensitive to the operator content of the CFT, hence it can be adopted to discern the TLL states. The geometry is illustrated in the inset of Fig.~\ref{fig:mutual}, where a chain with PBC is divided into four intervals with lengths $r_{A}$, $r_{B}$, $r_{C}$ and $r_{D}$. The mutual information of the intervals $A$ and $B$ is defined by
\begin{equation}
I_{A:B}=S_{A}+S_{B}-S_{A\cup B}.
\end{equation}
For the SU(2)$_{1}$ Wess-Zumino-Witten (WZW) theory, the mutual information is given by~\cite{Furukawa2009, Calabrese2009c}
\begin{equation}
I_{A:B}=\frac{c}{3}\ln\frac{\sin\big(\pi(r_{A}+r_{C})/L\big)\sin\big(\pi(r_{B}+r_{C})/L\big)}{\sin(\pi r_{C}/L)\sin(\pi r_{D}/L)}.
\end{equation}
We take $r_{A}=r_{B}=r$ and $r_{C}=r_{D}=L/2-r$ in the calculations, and thus expect that
\begin{equation}
I_{A:B}= -\frac{1}{3}\ln \cos^2\frac{\pi r}{L},
\end{equation}
which is plotted as the solid curve in Fig.~\ref{fig:mutual}. The ground state of the IDWM with PBC is calculated with the density-matrix renormalization group (DMRG) method. The mutual information $I_{A:B}$ is plotted in Fig.~\ref{fig:mutual}, and turns out to match the solid curve perfectly. Therefore, this provides further support to the SU(2)$_{1}$ WZW effective theory of the IDWM.

\subsection{FM transition on the domain wall}

\begin{figure}[!tb]
\centering
\includegraphics[width=0.5\textwidth]{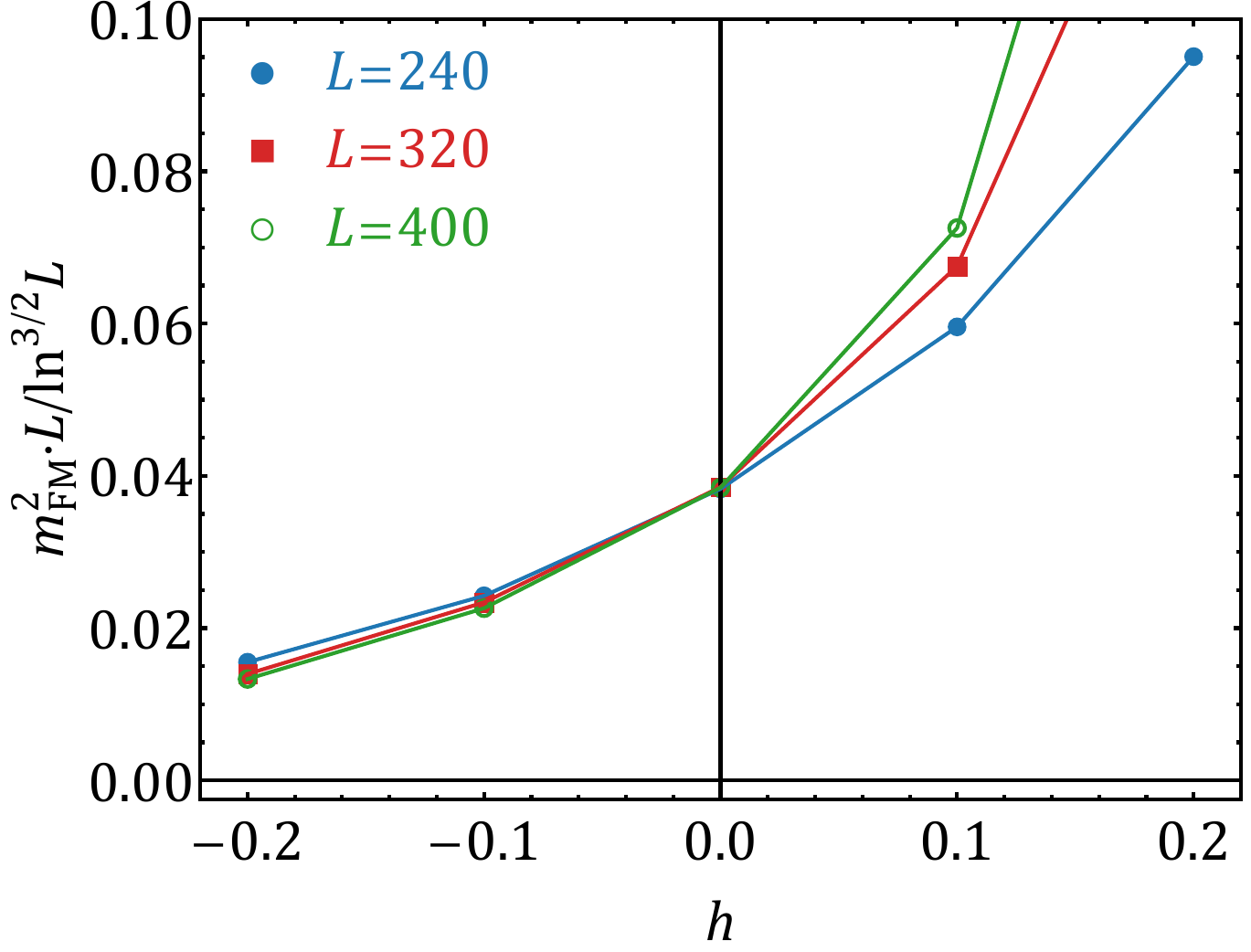}
\caption{The squared FM order parameter $m_{\mr{FM}}^{2}$ rescaled according to Eq.~(\ref{eq:mfm2}). The crossing of $m_{\mr{FM}}^{2}$ for different lattice sizes at $h=0$ confirms the SU(2)$_{1}$ WZW effective theory of the IDWM and supports the domain wall transition from the TLL phase to the FM order at $h=0$.}
\label{fig:cross}
\end{figure}

The RG analysis suggests that the FM transition of the perturbed domain wall model $H_{1}=H_{0}+H_{h}$ takes place at the IDWM at $h=0$, which is described by the SU(2)$_{1}$ WZW theory. As shown in Sec.~\ref{sec:mapping}, the AF order parameter in the Heisenberg chain representation maps to the FM order parameter of the domain wall model, hence the FM correlation function of the IDWM should obey the following scaling relation,
\begin{equation}
\langle \sigma_{l}^{z}\sigma_{l+r}^{z}\rangle\propto \frac{\ln^{1/2}r}{r},
\end{equation}
according to the AF correlation of the Heisenberg chain~\cite{Affleck1998a}, where the logarithmic factor comes from the marginally irrelevant perturbation. Therefore, we expect that the squared FM order parameter has the following finite-size scaling form at the QCP,
\begin{equation}
m_{\mr{FM}}^{2}=\frac{2}{L}\sum_{r=0}^{L/2}\langle \sigma_{L/2-r/2}^{z}\sigma_{L/2+r/2}^{z}\rangle\propto \frac{\ln^{3/2}L}{L}
\label{eq:mfm2}
\end{equation}
In Fig.~\ref{fig:cross}, numerical results of $m_{\mr{FM}}^{2}$ for different lattices sizes are rescaled according to Eq.~(\ref{eq:mfm2}), and the crossing at $h=0$ confirms the SU(2)$_{1}$ WZW effective theory of the IDWM, and supports the domain wall transition from the TLL phase to the FM order at $h=0$.

\subsection{Phase boundaries}

\begin{figure}[!tb]
\centering
\includegraphics[width=\textwidth]{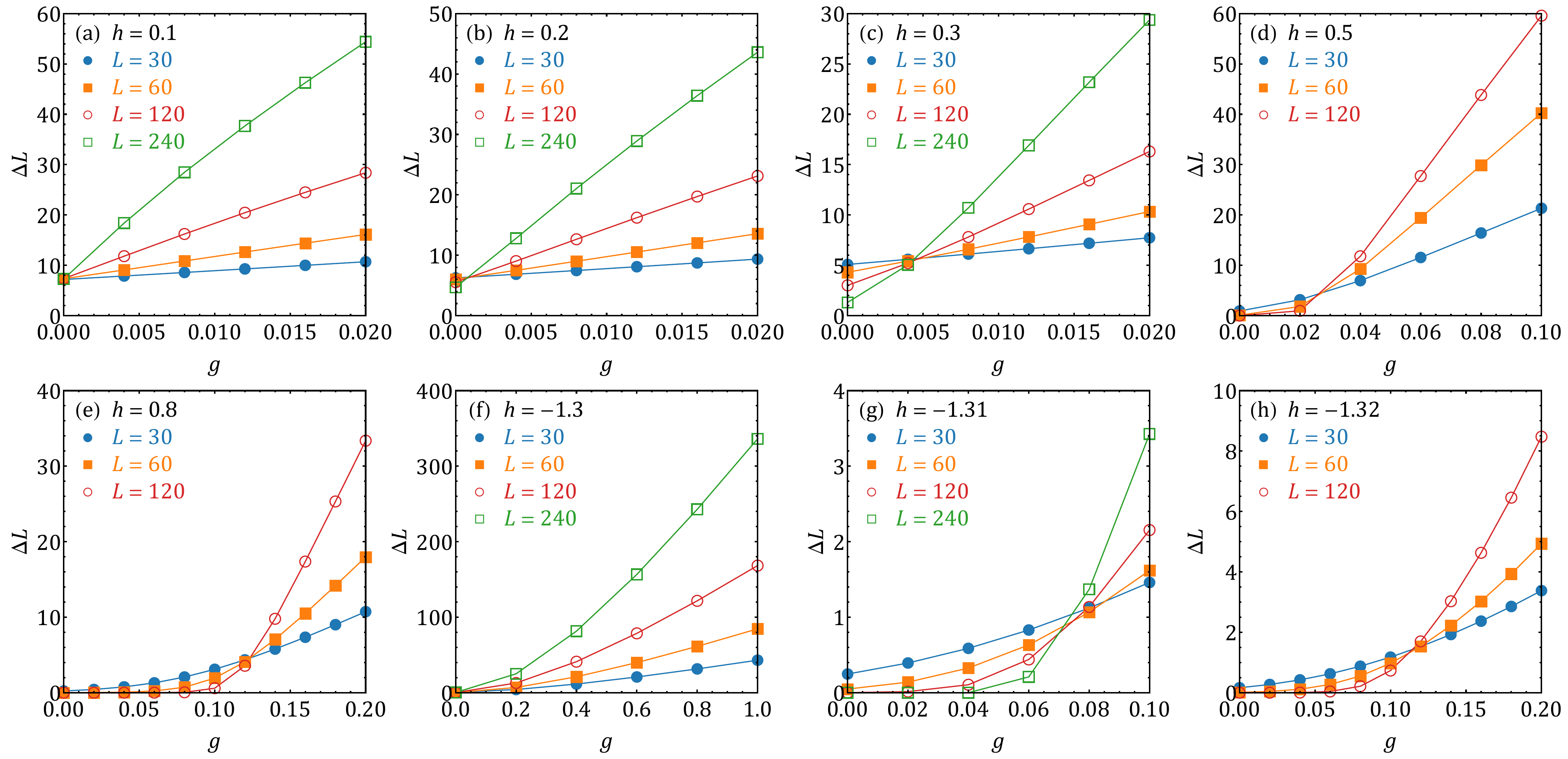}
\caption{The crossing of the quantity $\Delta L$ for different lattice sizes yields the estimate of phase boundaries of the generalized spin chain.}
\label{fig:gaps}
\end{figure}

The phase boundaries of the generalized spin chain are extracted with the crossing of the quantity $\Delta L^{z}$, where $\Delta$ is the energy gap obtained with the DMRG calculations, and $z$ is the dynamical exponent at the QCP, and we expect $z=1$ along the (1+1)D Ising transition lines of the generalized spin chain. $\Delta L^{z}$ is scale invariant at the QCP, thus its values obtained for different lattice sizes should cross at the QCP. The numerical results are plotted in Fig.~\ref{fig:gaps}.

\bibliography{C:/Documents/ResearchNotes/BibTex/library}